\documentclass[a4,12pt]{article}

\usepackage{amssymb,cancel,amsmath,indentfirst}
\usepackage{hyperref}

\begin{document}

\title{\bf Surface effect on the wrinkling of an elastic sheet under tension}%

\author{Jie Gu \\\emph{\small State Key Laboratory of Automotive Safety and Energy,}\\ \emph{ \small  Tsinghua University, Beijing 100084, People's Republic of China}}

\date{\small September 2013}%

\maketitle

\begin{abstract}
 Wrinkling of stretched elastic sheets is widely observed, and  the scaling relations between the amplitude
and wavelength of the wrinkles have been proposed by Cerda and Mahadevan \cite{Cerda2003}. However, the surface effects should be taken into account when the sheet is even thinner. The surface energy was considered in this work, and the discrepancies with the classical theory has been discussed. A dimensionless parameter has been proposed to represent the size-dependence. A method of characterizing mechanical properties of thin film using wrinkles considering surface effects has also been proposed.
\end{abstract}

\bigskip

Thin sheets are subject to bending, and wrinkles could be observed in daily life. In fact those wrinkles are involved in many facial expressions such as the skin of the forehead on a surprised face, or the stretching of a plastic bag. The mechanism of the latter is more counterintuitive, because thin sheets deform out of plane even under pure in-plane tension. Tension field theory could answer this question partly, and Cerda and Mahadevan complemented this classical theory \cite{Cerda2003, Cerda2002}. The sheet is seeking for a configuration which minimizes the total elastic energy and leads to the scaling relations between the amplitude and wavelength of the wrinkles. These relations have been used in many applications. An application of the wrinkle mechanics to scars and wound healing has been reported in Cerda \cite{Cerda2005}. Wrinkles produced by living-cells locomotion on an elastic membranes could be used to test forces generated by them \cite{Burton1999}. Wrinkles could also be used in mechanical characterization of thin elastic sheets such as cell \cite{Bernal2007}, graphene \cite{Bao2009}, polymer films \cite{Huang2007}, and manufacturing process with advanced materials \cite{Chan2006}. In a classical elasticity theory, the total energy could be divided into the part of stretching and bending for a thin sheet, while the surface energy should also be taken into account when the sheet is even thinner, since there are a lot of works reported, including many of those mentioned above, which have applied this result to \emph{ultra} thin sheets.

Surface effect could be prominent for a structure at small length scales, due to the high surface-to-volume ratio. Natural frequency of a nanotube or beam could be altered by surface effect \cite{Wang2007,Farshi2010}. Effective elastic properties of solids containing nano-inhomogeneities could be size-dependent with consideration of surface effects \cite{Sharma2003,Duan2005}. Instabilities could be at times either fatal or utilized in nano/microelectro mechanical systems (N/MEMS), and surface effects are also involved in these size-dependent phenomena \cite{Wang2009,Fu2011}. Moreover, the incorporation of surface effects eliminates the oscillatory singularities of the stress fields at the crack tips \cite{Kim2011}.  Many attempts have been made to make clear of and describe the behavior when dimensions are in the order of 100 nm, where the surface effect is usually not negligible. Calculations at the scale of atoms are useful \cite{Shenoy2005}, while a model in the sense of continuum is often more convenient and macroscopic behavior could be better revealed and studied using this kind of models. Among them the model proposed by Gurtin and Murdoch has been widely used \cite{Gurtin1975,Gurtin1978}. In their model, additional boundary conditions and surface constitutive relations are added to the classical equations in elasticity theory. The equations are summarized here.

In the bulk, the classical equations still hold:
\begin{equation}
  \begin{aligned}
 & \sigma _{ij,j}^B  = 0\\
 & \sigma _{ij}^B = {C_{ijkl}}{\varepsilon _{kl}}
   \end{aligned}
\end{equation}

There is also an equation of equilibrium on the surface, which resembles the Laplace's equation in fluid, and another equation resembles the constitutive equation in the bulk:
\begin{equation}
  \begin{aligned}
 & \left\langle {\sigma _{\beta \alpha }^B} \right\rangle {n_\beta } + \sigma _{\beta \alpha ,\beta }^S=0; \quad {\left \langle \sigma _{ji}^B \right \rangle}{n_j}{n_i} = \sigma _{\alpha \beta }^S{\kappa _{\alpha \beta }}\\
 & \sigma _{\beta \alpha }^S = {\sigma _0}{\delta _{\beta \alpha }} + 2\left( {{\mu _S} - {\sigma_0}} \right)\varepsilon _{\alpha \beta }^S + \left( {{\lambda _S} + {\sigma _0}} \right){\varepsilon _{\lambda \lambda }}{\delta _{\beta \alpha }}
   \end{aligned}
\end{equation}
In these equations, superscripts $B$ and $S$ represent bulk and surface respectively. The indices $i,j,k,l$, running from $1$ to $3$, are for three-dimensional bulk and $\alpha,\beta$ are for two-dimensional surface. $\lambda$ and $\mu$ are the Lam\'e constants for the bulk, $\lambda_S$ and $\mu_S$ for the surface, and $\sigma_0$ is the residual surface tension.

Here we considered the case studied by Cerda and Mahadevan\cite{Cerda2003}, and added the term of surface energy to make it valid for an ultrathin sheet. A thin isotropic elastic sheet with thickness $t$, width $W$, length $L$, Young's modulus $E$ and poison ratio $\nu$ suffers a tension of $T$, and  is stretched to a longitudinal strain $\gamma$. When $\gamma  > {\gamma _C}$, a critical stretching strain, the sheet will no longer be flat and wrinkles are formed. The out-of-plane displacement is represented by $\zeta$ . Therefore, ${\varepsilon _{xx}} = \frac{1}{2}\zeta _x^2 + {u_x} - z{\zeta _{xx}}$. Assumptions are that there is no shear stress in the sheet, and $\frac12\zeta _y^2 + {v_y}$ could be ignored so ${\varepsilon _{yy}} = - z{\zeta _{yy}}$.

The functional to be minimized should be
\begin{equation}
U = {U_B} + {U_S} + {U_{\rm Surface}} + V - {\cal L}
\end{equation}
${U_B} = \frac{1}{2}\int_A {B\zeta _{yy}^2dA}$ is the bending energy, where ${\zeta _{xx}}$ was neglected, and $B = Et^3/ \left[ 12 \left( 1-\nu^2 \right) \right] $  is the bending stiffness. ${U_S} = \int_A {\frac{{Et}}{2}{{\left( {\frac{1}{2}\zeta _x^2 + {u_x}} \right)}^2}dA} $
  is the stretching energy.

Surface energy of Gurtin-Murdoch model has been discussed before \cite{Ru2010}, and we choose the form:
  \begin{equation}
  \begin{aligned}
  & {U_{\rm Surface}} = \int_0^l {\int_s { \psi dsdl} } \\
 & = \int_0^l {\int_s {[{\sigma _0}\left( {1 + \varepsilon _{\alpha \alpha }^S} \right) + \frac{1}{2}{\lambda _S}{{\left( {\varepsilon _{\alpha \alpha }^S} \right)}^2}} }
 + {\mu _S}\varepsilon _{\alpha \beta }^S\varepsilon _{\alpha \beta }^S]dsdl\\
 & =U_{1}+U_{2}+U_{3}
  \end{aligned}
  \end{equation}
  which we added here to take surface effect into consideration.

  The integral is taken over all the surface. Since $t \ll W$, only the two surfaces corresponding to the long sides are to be considered. At the surface, $z =  \pm t/ 2$, so terms linear in $z$ will cancel each other. We finally get the three terms of the surface energy:
\begin{subequations}
\begin{equation}
U_1  = 2{\sigma _0}\int_A {\left( {1 + {u_x} + \frac{1}{2}\zeta _x^2} \right)dA}
\end{equation}
\begin{equation}\label{lambda}
{U_{2}} = {\lambda _S}\int_A {\left( {\frac{1}{4}\zeta _x^4 + u_x^2 + \frac{{{t^2}}}{4}\zeta _{xx}^2 + {u_x}\zeta _x^2 + \frac{{{t^2}}}{4}\zeta _{yy}^2} \right)dA}
\end{equation}
\begin{equation}\label{mu}
{U_3} = 2{\mu _S}\int_A {\left( {\frac{1}{4}\zeta _x^4 + u_x^2 + \frac{{{t^2}}}{4}\zeta _{xx}^2 + {u_x}\zeta _x^2 + \frac{{{t^2}}}{4}\zeta _{yy}^2} \right)dA}
\end{equation}
\end{subequations}

The observation of the equations \eqref{lambda} and \eqref{mu} leads to the definition of a new parameter called effective surface Lam\'e constants, which is combination of the two surface parameters $\kappa_S \equiv \lambda_S+2\mu_S$. Due to the condition of inextensibility, $\int_0^W {\left( {\frac{1}{2}\zeta _y^2 + {v_y}} \right)} dy = 0$, and use it as a constraint, ${\cal L} = \int_A {b(x)\left( {\frac{1}{2}\zeta _y^2 + {v_y}} \right)} dA$. $V$ is the potential of the tension, and is given by $V=-\int_A {T u_x dA}$.

 The minimization indicates that the variation vanishes, i.e. ${\delta U} /{\delta \zeta }= 0$ and $\delta U/ \delta u_x =0$,   and finally yields two equations:
\begin{equation}\label{B}
  \begin{aligned}
B'{\zeta _{yyyy}} - \left[ {Et\left( {\frac{1}{2}\zeta _x^2 + {u_x}} \right) + 2{\sigma _0} - 2\kappa_S {u_x}} \right]{\zeta _{xx}}\\
 + \kappa_S \left( { - 3\zeta _x^2{\zeta _{xx}} + \frac{{{t^2}}}{2}{\zeta _{xxxx}}} \right) + b(x){\zeta _{yy}} = 0
   \end{aligned}
\end{equation}
\begin{equation}\label{ux}
Et\left( {\frac{1}{2}\zeta _x^2 + {u_x}} \right) + 2{\sigma _0} + 2\kappa_S \left( {\frac{1}{2}\zeta _x^2 + {u_x}} \right) - T = 0
\end{equation}
where $B'\equiv B + \frac{{{t^2}}}{2}\kappa_S$.

The equation \eqref{ux} is actually the equation of equilibrium for a sheet under the tension, and holds no matter whether the wrinkles are formed or not. The stretching stiffness $E$ has been changed by the surface properties, so the effective stiffness could be denoted as $E+2\kappa_S/t$, indicating that surface effects grows when thickness is small. $ -2{\sigma _0}$ behaves like a tension, and could be superposed over tension $T$. A simple calculation reveals the longitudinal strain and put it in \eqref{B} which controls the wrinkles. Assuming $\frac12 \zeta_x^2 \ll u_x$ yields:
\begin{equation}
B'{\zeta _{yyyy}} - \left[ {\left( {T - 2{\sigma _0}} \right)\theta  + 2{\sigma _0}} \right]{\zeta _{xx}} + b{\zeta _{yy}}
 + \kappa_S \left( { - 3\zeta _x^2{\zeta _{xx}} + \frac{{{t^2}}}{2}{\zeta _{xxxx}}} \right) = 0
 \end{equation}
where $\theta  = \frac{{Et - 2\kappa_S}}{{Et + 2\kappa_S}}$. Let's introduce a dimensionless parameter $\rho \equiv \kappa_S/Et$, which combines the stretching stiffness $E$ of bulk, the thickness $t$ and the effective surface Lam\'e constants $\kappa_S$. This parameter is in fact a measure of size-dependence. On the other hand, we could also define an intrinsic thickness $t_0 \equiv \kappa_s/E$. When $t \sim t_0$ or $\rho \sim 1$, surface effects are considerable. Use $\rho$ then $\theta=\frac{1-2\rho}{1+2\rho}$.

Use the dimensionless coordinates $\tilde x=x/L$, $\tilde y=y/W$ and $\tilde \zeta=\zeta/t$, and consider the relation $t \ll W \ll L$, we could find that the coefficient of the last term is much smaller than the others and could be ignored. Finally the equation reads:
\begin{equation}
B'{\zeta _{yyyy}} - \left[ {\left( {T - 2{\sigma _0}} \right)\theta  + 2{\sigma _0}} \right]{\zeta _{xx}} + b{\zeta _{yy}} = 0
\end{equation}

Compared with the corresponding equation in Cerda and Mahadevan's work \cite{Cerda2003}, the results could be regarded as being acquired with the following replacements:
\begin{equation}\label{replace}
\begin{aligned}
& B'=B + \frac{{{t^2}}}{2}\kappa_S  \to B \\
& T' \equiv \left( T - 2{\sigma _0} \right) \theta + 2\sigma_0 \to T
\end{aligned}
\end{equation}

Therefore, $B'=B + \frac{{{t^2}}}{2}\kappa_S$  is the ¡°effective¡± bending stiffness, and $\left( T - 2{\sigma _0} \right) \theta + 2\sigma_0 $ is the effective tension.

Consider the asymptotic behavior first. On one hand, the deviation of bending stiffness caused by surface effect is
\begin{equation}
\frac{{\Delta B}}{B} = \frac{{B' - B}}{B} = \left[ {6\left( {1 - {\nu ^2}} \right)} \right] \rho.
\end{equation}

This relative discrepancy is inversely proportional to thickness $t$, and will increase dramatically when  $t$ gets smaller. In other words, the difference will be more significant when the sheet is thinner. Moreover, the sign¡ªwhether it¡¯s stiffer or more compliant¡ªis supposed to be determined by the sign of the material parameters $\lambda_S$  and $\mu_S$.

On the other hand, when $Et \gg \kappa_S$, $\Delta B/B \to 0$ and $\theta \to 1 $, it reduced to the case in the classical case \cite{Cerda2003}. In other words, when the sheet is thick enough, or the material parameters of the surface are small enough, it described the macroscopic properties, because the surface effect is not significant at all. There could be several analogies to other physical systems. In optics, small wavelengths lead to geometrical optics from wave optics. In quantum mechanics, Planck's constant determines the extent of quantization of the system. The intrinsic thickness $t_0$ in our case is the most similar to the thermal wavelength $\lambda$ of gas, which is also the property of the system itself. When $\lambda$ is much smaller than the average interparticle distance, the system could be described classically; when $\lambda$ gets larger, quantum effects become more significant.

Let's see how surface energy affects the stretching of the sheet. If $\kappa_S$ is small compared with $Et$, as we have discussed above, \eqref{B}, which is independent of $\sigma_0$, reduced to the one in the original case. However, $ -2{\sigma _0}$ behaves like a tension, which could be superposed on tension $T$, and affects the longitudinal strain. Since the critical strain $\gamma_C$, above which the wrinkles are formed, would be independent on $\sigma_0$, the initiation of the wrinkling depends on the residual stress. It could be harder for the sheet to wrinkle if $\sigma_0>0$.

With the differential equation of $\zeta$, the scaling relations of the amplitude and wavelength of the wrinkles are to be determined, with minor modifications from Cerda's paper. We briefly recall the derivations here.

Separation of variables is made so $\zeta=\Sigma e^{ik_n y}X_n\left(x\right)$. It yields an equation $\frac{d^2 X_n}{dx^2}+\omega^2_n X_n=0$, where $\omega_n^2=\left( bk_n^2 -B' k_n^4\right)/T'$. The solution with the least bending energy is $\zeta=A \sin(\pi x/L) \cos (ky+\phi)$. The condition of inextensibility yields $A^2 k^2 W/8 \approx \Delta$, where $\Delta$ is the compressive transverse width reduction of the sheet. Therefore the total energy $U = B'{k^2}L\Delta  + \frac{{{\pi ^2}T\Delta }}{{{k^2}L}} + \rm const.$. The terms with respect to $\zeta_x^4, \zeta_{xx}^2$ have been ignored. An important and a little surprising observation is that even though $B$ has been changed to $B'$ here, $T$ returned. Minimizing $U$ yields
\begin{equation}\label{result}
\lambda  = 2\sqrt \pi  {\left( {\frac{{B'}}{T}} \right)^{\frac{1}{4}}}{L^{\frac{1}{2}}}, \quad A = \frac{{\sqrt 2 }}{\pi }{\left( {\frac{\Delta }{W}} \right)^{\frac{1}{2}}}\lambda.
\end{equation}

If wrinkles are used to characterize a thin solid film, it is an inverse problem of what is considered here, since the amplitude $A$ and wavelength $\lambda$ are measured and Young's modulus $E$ is what is to be calculated and determined. We first determine at what thickness the surface effect could not be ignored. From \eqref{result} the deviation of $E$'s with or without surface effects considered is in fact that of $B$'s, which has been presented in \eqref{replace}. The following data taken from Gurtin and Murdoch's work \cite{Gurtin1978} is used here as an simple example to illustrate this deviation. $E=5.625 \times 10^{10}\ \rm N/m^2,\, \nu=0.25, \, \lambda_S=7 \times 10^3\ \rm N/m, \, \mu_S=8\times 10^3\ \rm N/m, \, \sigma_0=110\ \rm N/m$. A simple calculation leads to $t_0=409 \ \rm nm$. When thickness $t \sim 100\ \rm nm$, surface effects are considerable.

We conclude by proposing a minor modified method of characterizing mechanical properties of thin film using wrinkles with surface effects considered. Using the inverse relation of \eqref{result} we could get $B'$. However, Young's modulus could not be deduced directly from $B'$ as in a classical isotropic thin plate. Consider the definition of $B'$, so opbviously $B'/t^2= Et/ \left[ 12 \left( 1-\nu^2 \right) \right]+ \kappa_S /2$. In other words, $B/t^2 \sim t$ is a curve through origin in the classical theory, while surface effects offset the curve with linear form preserved. The slope corresponds to $E$ and the intercept corresponds to $\kappa_S$. Fitting curves to several experimental data points indicates both paramters.


\begin{thebibliography}{10}

\bibitem{Cerda2003}
E. Cerda and L. Mahadevan,
\newblock Phys. Rev. Lett. 90 (2003) 074302.

\bibitem{Cerda2002}
E. Cerda, K. Ravi-Chandar and L. Mahadevan,
\newblock Nature 419 (2002) 579.

\bibitem{Cerda2005}
E. Cerda,
\newblock J. Biomech. 38 (2005) 1598.

\bibitem{Burton1999}
K. Burton, J.H. Park and D.L. Taylor,
\newblock Mol. Biol. Cell 10 (1999) 3745.

\bibitem{Bernal2007}
R. Bernal et~al.,
\newblock Appl. Phys. Lett. 90 (2007) 063903.

\bibitem{Bao2009}
W. Bao et~al.,
\newblock Nature nanotechnology 4 (2009) 562.

\bibitem{Huang2007}
J. Huang et~al.,
\newblock Science 317 (2007) 650.

\bibitem{Chan2006}
E.P. Chan and A.J. Crosby,
\newblock Advanced Materials 18 (2006) 3238.

\bibitem{Wang2007}
G.F. Wang and X.Q. Feng,
\newblock Appl. Phys. Lett. 90 (2007) 231904.

\bibitem{Farshi2010}
B. Farshi, A. Assadi and A. Alinia-Ziazi,
\newblock Appl. Phys. Lett. 96 (2010) 093105.

\bibitem{Sharma2003}
P. Sharma, S. Ganti and N. Bhate,
\newblock Appl. Phys. Lett. 82 (2003) 535.

\bibitem{Duan2005}
H. Duan et~al.,
\newblock Journal of the Mechanics and Physics of Solids 53 (2005) 1574.

\bibitem{Wang2009}
G.F. Wang and X.Q. Feng,
\newblock Appl. Phys. Lett. 94 (2009) 141913.

\bibitem{Fu2011}
Y. Fu and J. Zhang,
\newblock Applied Mathematical Modelling 35 (2011) 941.

\bibitem{Kim2011}
C.I. Kim, P. Schiavone and C.Q. Ru,
\newblock Proceedings of the Royal Society A: Mathematical, Physical and
  Engineering Science 467 (2011) 3530.

\bibitem{Shenoy2005}
V.B. Shenoy,
\newblock Physical Review B 71 (2005) 094104.

\bibitem{Gurtin1975}
M.E. Gurtin and A.I. Murdoch,
\newblock Archive for Rational Mechanics and Analysis 57 (1975) 291.

\bibitem{Gurtin1978}
M.E. Gurtin and A. Ian~Murdoch,
\newblock International Journal of Solids and Structures 14 (1978) 431.

\bibitem{Ru2010}
C. Ru,
\newblock Science China Physics, Mechanics and Astronomy 53 (2010) 536.

\end{thebibliography}

\end{document}